\documentclass{article}
\usepackage{graphicx}
\usepackage{amsmath}

\input{tcilatex}

\begin{document}

\begin{center}
\textbf{THE LONGITUDINAL DYNAMIC CORRELATION AND DYNAMIC SUSCEPTIBILITY OF
THE ISOTROPIC XY-MODEL ON THE 1D ALTERNATING SUPERLATTICE}$^{\S }$

\bigskip

J. P. de Lima

Departamento de F\'{i}sica da Universidade Federal do Piau\'{i}

Campus Ministro Petr\^{o}nio Portela, 64049-550 Teresina, Piau\'{i}, Brazil

and

L. L. Gon\c{c}alves*

Departamento de F\'{i}sica da Universidade Federal do Cear\'{a}

Campus do Pici, C.P. 6030, 60451-970 Fortaleza, Cear\'{a}, Brazil

\bigskip

\textbf{Abstract}
\end{center}

The dynamic susceptibility $\chi _{Q}^{zz}(\omega )$ of \ the isotropic
XY-model (s=1/2) on the alternating superlattice (closed chain) in a
transverse field $h$ is obtained exactly at arbitrary temperatures. It is
determined from the results obtained for the dynamic correlations $%
<S_{jn}^{z}(t)S_{lm}^{z}(0)>,$ which have been calculated by introducing the
generalized Jordan-Wigner transformation, by using Wick's theorem and by
reducing the problem to a diagonalization of a finite matrix. The static
properties are also reobtained within this new formalism and all exact
results are determined for arbitrary temperatures. Explicit results are
obtained numerically in the limit $T=0,$ where the critical behaviour
occurs. A detailed analysis is presented for the behaviour of the static
susceptibility $\chi _{Q}^{zz}(0)$, as a function of the transverse field $h$%
, and for the frequency dependency of the dynamic susceptibility $\chi
_{Q}^{zz}(\omega )$. It is also shown, in this temperature limit, that
within the magnetization plateaus which correspond to the different phases,
even when the induced magnetization is not saturated, the effective dynamic
correlation, $<\sum\limits_{n;m\in cell:\text{ }%
j;l}S_{jn}^{z}(t)S_{lm}^{z}(0)>$ , is time independent, which constitutes an
unexpected result.

\bigskip

$^{\S }$ A preliminary version of this work has been presented at STATPHYS
21,

\ July 2001, Cancun, Mexico.

*Corresponding author.

\textit{E-mail address: }lindberg@fisica.ufc.br\textit{\ }\ (L. L. Gon\c{c}%
alves)

PACS: 05.90.+m;75.10Jm

Keywords: XY model; dynamic susceptibility; dynamic correlation;

\ \ \ \ \ \ \ \ \ \ \ \ \ \ \ \ \ one-dimensional superlattice.

\bigskip \newpage

\bigskip {\Large 1. Introduction}

The homogeneous one-dimensional XY-model introduced by Lieb et al.[1] has
been object of many studies. Its importance comes mainly from the fact that
it is among the few many-body problems which can be solved exactly. Although
the static and dynamic properties of the general homogenous model have been
studied since its introduction (see e.g. Ref. [2] and references therein),
the study of the inhomogeneous periodic model has been restricted to the
alternating chain [3-5]. Only recently have more general inhomogeneous
models been object of study , namely, the alternating superlattice [6-8] and
\ the inhomogeneous periodic model composed of inhomogeneous segments of
finite length [9,10]. The results of these papers have been restricted to
the static properties of the isotropic model and up to this date, to the
best of our knowledge, no results have been obtained for the dynamic
properties of these models.

In this paper we study the dynamic properties of the isotropic model on the
alternating \ one-dimensional\ superlattice. The dynamic properties are
restricted to the longitudinal direction and they correspond to an extension
of our previous results [6-8]. In section 2 we present the solution of the
model and the basic results. In section 3 we obtain the dynamic correlations
in the field direction and in section 4 the longitudinal dynamic
susceptibility. Finally in section 5 we present the results and the main
conclusions.

\bigskip

{\Large 2. The model: basic results}

We consider the isotropic XY model on the one-dimensional alternating
superlattice$,$ which we have been able to solve exactly[7]. The
superlattice consists of N\ cells, composed of two subcells, A and B, with $%
n_{A}$ and $n_{B}$ sites, respectively. The\textit{\ l-th} unit cell is
shown in Fig.1, and the distance $s$ between two consecutive sites is taken
equal to one.

If we assume periodic boundary conditions for a chain with \textit{N} cells,
the Hamiltonian of the \textit{XY-}model [7] can be written in the form

\bigskip

\begin{eqnarray}
\ H &=&-\frac{1}{2}\dsum\limits_{l=1}^{N}\left\{ \left[ \dsum%
\limits_{m=1}^{n_{A}-1}J_{A}S_{l,m}^{A^{+}}S_{l,m+1}^{A^{-}}+\dsum%
\limits_{m=1}^{n_{B}-1}J_{B}S_{l,m}^{B^{+}}S_{l,m+1}^{B^{-}}\right. \right. 
\notag \\
&&+\left. J\left(
S_{l,n_{A}}^{A^{+}}S_{l,1}^{B^{-}}+S_{l,n_{B}}^{B^{+}}S_{l+1,1}^{A^{-}}%
\right) +h.c.\right]  \notag \\
&&\ \left.
+\dsum\limits_{m=1}^{n_{A}}2h_{A}S_{l,m}^{A^{Z}}+\dsum%
\limits_{m=1}^{n_{B}}2h_{B}S_{l,m}^{B^{Z}}\right\} ,  \TCItag{1}
\end{eqnarray}

\bigskip

\noindent where $l$ identifies the cell, $S^{\pm }=S^{x}\pm iS^{y}$, $J$ is
the exchange parameter between spins at the interfaces, $J_{A}(J_{B})$ the
exchange parameter between spins within the subcell $A(B),$ and $%
h_{A}(h_{B}) $ is the transverse field within the subcell $A(B).$ The spin
operators can be expressed in terms of fermion operators using the
generalized Jordan-Wigner transformation [6]

\bigskip

\begin{equation}
S_{l,m}^{A^{+}}=\exp \left\{ i\pi
\dsum\limits_{n=1}^{l-1}\dsum\limits_{r=1}^{n_{A}}a_{n,r}^{\dag
}a_{n,r}+i\pi \dsum\limits_{r=1}^{m-1}a_{l,r}^{\dag }a_{l,r}+i\pi
\dsum\limits_{n=1}^{l-1}\dsum\limits_{r=1}^{n_{B}}b_{n,r}^{\dag
}b_{n,r}\right\} a_{l,m}^{\dag }\ ,  \tag{2}
\end{equation}

\bigskip

\begin{equation}
S_{l,m}^{B^{+}}=\exp \left\{ i\pi
\dsum\limits_{n=1}^{l}\dsum\limits_{r=1}^{n_{A}}a_{n,r}^{\dag }a_{n,r}+i\pi
\dsum\limits_{r=1}^{m-1}b_{l,r}^{\dag }b_{l,r}+i\pi
\dsum\limits_{n=1}^{l-1}\dsum\limits_{r=1}^{n_{B}}b_{n,r}^{\dag
}b_{n,r}\right\} b_{l,m}^{\dag }\ ,  \tag{3}
\end{equation}

\bigskip

\noindent and, by introducing this transformation, the Hamiltonian can be
written in the form

\bigskip

\begin{eqnarray}
\ H &=&-\frac{1}{2}\dsum\limits_{l=1}^{N}\left\{
\dsum\limits_{m=1}^{n_{A}-1}J_{A}a_{l,m}^{\dag
}a_{l,m+1}+\dsum\limits_{m=1}^{n_{B}-1}J_{B}b_{l,m}^{\dag }b_{l,m+1}\
+\right.  \notag \\[0.01in]
&&+J\left( a_{l,n_{A}}^{\dag }b_{l,1}+b_{l,n_{B}}^{\dag }a_{l+1,1}\right)
+h.c.  \notag \\
&&\ \left. +\dsum\limits_{m=1}^{n_{A}}2h_{A}\left( a_{l,m}^{\dag
}a_{l,m}-1\right) +\dsum\limits_{m=1}^{n_{B}}2h_{B}\left( b_{l,m}^{\dag
}b_{l,m}-1\right) \right\} +\Phi ,  \TCItag{4}
\end{eqnarray}

\bigskip

\noindent where $a^{\prime }$s and $b^{\prime }$s are fermion operators, and 
$\Phi ,$ given by

\bigskip

\begin{equation}
\Phi =\frac{J}{2}\left( b_{N,n_{B}}^{\dag }a_{1,1}+h.c.\right) \exp \left[
i\pi \left( \dsum\limits_{l=1}^{N}\dsum\limits_{r=1}^{n_{A}}a_{l,r}^{\dag
}a_{l,r}+\dsum\limits_{l=1}^{N}\dsum\limits_{r=1}^{n_{B}}b_{l,r}^{\dag
}b_{l,r}\right) \right] \qquad ,  \tag{5}
\end{equation}

\bigskip

\noindent is a boundary term which will be neglected. As it has been shown
[11], this boundary term, in the thermodynamic limit, does not affect the
excitation spectrum, the static properties of the system, nor the dynamic
correlation function in the field direction. Introducing the Fourier
transforms [8],

\bigskip

\begin{eqnarray}
A_{Q,m} &=&\frac{1}{\sqrt{N}}\dsum\limits_{l}\exp \left( -iQdl\right)
a_{l,m}\qquad ,  \TCItag{6} \\
&&  \notag \\
B_{Q,m} &=&\frac{1}{\sqrt{N}}\dsum\limits_{l}\exp \left( -iQdl\right)
b_{l,m}\qquad ,  \TCItag{7}
\end{eqnarray}

\bigskip

\noindent where $Q=2\pi n/Nd,$ $n=1,2,...,N,$ and $d=n_{A}+n_{B}$ is the
size of the cell, the Hamiltonian can be written as

\bigskip

\begin{equation}
H=\dsum\limits_{Q}H_{Q}+N\left( \frac{n_{A}h_{A}+n_{B}h_{B}}{2}\right)
\qquad ,  \tag{8}
\end{equation}

\noindent where $H_{Q}$ is given by

\bigskip

\begin{eqnarray}
H_{Q} &=&-\sum_{m=1}^{n_{A}}h_{A}A_{Q,m}^{\dag
}A_{Q,m}-\sum_{m=1}^{n_{B}}h_{B}B_{Q,m}^{\dag }B_{Q,m}-  \notag \\
&&\sum_{m=1}^{n_{A}-1}\left( \frac{J_{A}}{2}A_{Q,m}^{\dag
}A_{Q,m+1}+h.c\right) -\sum_{m=1}^{n_{B}-1}\left( \frac{J_{B}}{2}%
B_{Q,m}^{\dag }B_{Q,m+1}+h.c\right) -  \notag \\
&&\frac{J}{2}\left[ A_{Q,n_{A}}^{\dag }B_{Q,1}+B_{Q,n_{B}}^{\dag
}A_{Q,1}e^{-iQd}+h.c.\right] +N\frac{\left( n_{A}h_{A}+n_{B}h_{B}\right) }{2}%
.  \TCItag{9}
\end{eqnarray}

\bigskip

\noindent We can also write $H$ in the form

\bigskip

\begin{equation}
H=\sum_{Q}V_{Q}^{\dag }\mathbf{T}(Q)V_{Q},  \tag{10}
\end{equation}

\bigskip

where

\bigskip

\begin{equation}
V_{Q}^{\dag }=\left( A_{Q,1}^{\dag },...,A_{Q,n_{A}}^{\dag },B_{Q,1}^{\dag
},...,B_{Q,n_{B}}^{\dag }\right)  \tag{11}
\end{equation}

\bigskip

\noindent and $\mathbf{T}(Q)$ is a matrix of dimension $d\times d$ which
represents the quadratic Hamiltonian $H_{Q}.$ Since $[H_{Q,}H_{Q^{\prime
}}]=0,$ the diagonalization of the Hamiltonian is reduced to the
diagonalization of $H_{Q}$, which can be written as a free fermion system

\bigskip

\begin{equation}
H_{Q}=\sum_{k}E_{Qk}\xi _{Q,k}^{\dag }\xi _{Q,k},  \tag{12}
\end{equation}

\bigskip

\noindent where $E_{Qk}$ are the diagonal elements of $\ U_{Q}\mathbf{T}%
_{Q}U_{Q}^{\dag },$ and $U_{Q}$ is the unitary transformation which
diagonalizes $H_{Q},$which is determined numerically. The energy spectrum
presents $d$ branches and, by using the unitary transformation $U_{Q},$ we
can express the fermion operators $a^{\prime }s$ and $b^{\prime }s$ in terms
of the $\xi ^{\prime }s$, as

\bigskip

\begin{eqnarray}
A_{Q,m} &=&\dsum\limits_{k}u_{Q,m,k}\xi _{Q,k},  \TCItag{13} \\
B_{Q,m} &=&\dsum\limits_{k}u_{Q,n_{A}+m,k}\xi _{Q,k},  \TCItag{14} \\
a_{l,m} &=&\frac{1}{\sqrt{N}}\dsum\limits_{Q,k}\exp \left( iQdl\right)
u_{Q,m,k}\xi _{Q,k},  \TCItag{15} \\
b_{l,m} &=&\frac{1}{\sqrt{N}}\dsum\limits_{Q,k}\exp \left( iQdl\right)
u_{Q,n_{A}+m,k}\xi _{Q,k},  \TCItag{16}
\end{eqnarray}

\bigskip

\noindent where $u_{Q,k,j}$ are defined as

\bigskip

\begin{equation}
\mathbf{T}_{Q}\left[ 
\begin{array}{c}
u_{Q,k,1} \\ 
... \\ 
u_{Q,k,n_{A}} \\ 
u_{Q,k,n_{A}+1} \\ 
... \\ 
u_{Q,k,n_{A}+n_{B}}
\end{array}
\right] =E_{Qk}\left[ 
\begin{array}{c}
u_{Q,k,1} \\ 
... \\ 
u_{Q,k,n_{A}} \\ 
u_{Q,k,n_{A}+1} \\ 
... \\ 
u_{Q,k,n_{A}+n_{B}}
\end{array}
\right] .  \tag{17}
\end{equation}

\bigskip

The general solution of this equation is determined numerically, although
analytical solutions can be found for some special cases. For instance for $%
n_{A}=n_{B}=2,h_{A}=h_{B}=h,$ we can express explicitly the solution in the
form

\bigskip

\begin{equation}
\omega _{Q}=-h\pm \frac{1}{\sqrt{2}}\sqrt{c\pm \sqrt{g\left( Qd\right) }}%
\qquad ,  \tag{18}
\end{equation}

\bigskip

\noindent where

\bigskip

\begin{eqnarray}
c &\equiv &\frac{J^{2}}{2}+\frac{1}{4}\left( J_{A}^{2}+J_{B}^{2}\right)
\qquad ,  \TCItag{19} \\
&&  \notag \\
g\left( Qd\right) &\equiv &\frac{1}{2}J^{2}J_{A}J_{B}\cos \left( Qd\right) +%
\frac{J^{2}}{4}\left( J_{A}^{2}+J_{B}^{2}\right) +\frac{1}{16}\left(
J_{A}^{2}-J_{B}^{2}\right) ^{2}\quad .  \TCItag{20}
\end{eqnarray}

\bigskip

As it has already been shown [7], the effect of a homogeneous field, $%
h_{A}=h_{B}=h,$ is to shift the zero field spectrum. For the special case
shown above this can also be shown directly from eq.(18) and, consequently,
the existence of a mode of zero energy will depend on the strength of the
field.

In passing, we would like to note that the spectrum can also be calculated
exactly by using the position space renormalization group approach [12], and
approximately by using a transfer matrix method [13]. Although the latter is
an approximate method, we have shown that it reproduces numerically the
exact result.

Within the formalism introduced above we can obtain easily all the
thermodynamic properties of the system. Then, we can express the internal
energy $U$ in the form

\bigskip

\begin{equation}
U=\sum\limits_{Q,k}E_{Qk}\left\langle n_{Qk}\right\rangle ,\text{ with\ \ \ }%
\left\langle n_{Qk}\right\rangle =\frac{1}{1+e^{\beta E_{Qk}}},  \tag{21}
\end{equation}

\bigskip

\noindent where $\beta =1/k_{B}T,$ $\ k_{B}$ is the Boltzmann constant and $%
T $, the absolute temperature, and from this expression we obtain
immediately the specific heat $C$

\bigskip

\begin{equation}
C=\frac{1}{k_{B}T^{2}}\sum\limits_{Q,k}E_{Qk}e^{\beta E_{Qk}}\left\langle
n_{Qk}\right\rangle ^{2}.  \tag{22}
\end{equation}

\bigskip

The induced magnetization per site and cell, which is defined as [7]$,$

\bigskip

\begin{equation}
\left\langle S_{cel}^{z}\right\rangle =\frac{1}{N\left( n_{A}+n_{B}\right) }%
\left[ \dsum\limits_{l=1}^{N}\left( \dsum\limits_{m=1}^{n_{A}}\left\langle
S_{l,m}^{A^{z}}\right\rangle +\dsum\limits_{m=1}^{n_{B}}\left\langle
S_{l,m}^{B^{z}}\right\rangle \right) \right] ,  \tag{23}
\end{equation}

\bigskip

can also be written in the form

\bigskip

\begin{equation}
\left\langle S_{cel}^{z}\right\rangle =\frac{1}{N\left( n_{A}+n_{B}\right) }%
\left[ \dsum\limits_{Qk}\left\langle n_{Qk}\right\rangle \right] -\frac{1}{2}%
,  \tag{24}
\end{equation}

\bigskip

and from this expression, by making the identification $h_{A}=h_{B}\equiv h$%
, we can obtain the isothermal susceptibility, $\chi _{T}^{zz}=\frac{%
\partial }{\partial h}\left\langle S_{cel}^{z}\right\rangle ,$ which is
given by

\bigskip

\begin{equation}
\chi _{T}^{zz}=\frac{\beta }{N\left( n_{A}+n_{B}\right) }\dsum\limits_{Qk}%
\left\langle n_{Qk}\right\rangle ^{2}\exp (\beta E_{Qk}).  \tag{25}
\end{equation}

\bigskip

{\Large 3. The Dynamic Two-Spin Correlation Function in the Field Direction}

The dynamic two-spin correlation function, $\left\langle
S_{j,m}^{A^{z}}(t)S_{j+R,n}^{A^{z}}(0)\right\rangle ,$ can be easily
obtained by writing explicitly the time-evolution of the operator $%
S_{j,m}^{A^{Z}}(t)$. This is obtained by using eqs. (2) and (15), which gives

\bigskip

\begin{eqnarray}
S_{j,m}^{A^{Z}}(t) &=&\frac{1}{N}\dsum\limits_{Q,k,Q^{\prime },k^{\prime
}}[\exp \left( -i(Q-Q^{\prime })dj\right) \exp (i(E_{Qk}-E_{Q^{\prime
}k^{\prime }})t)  \notag \\
&&u_{Q,k,m}^{\ast }u_{Q^{\prime },k^{\prime },m}\xi _{Q,k}^{\dag }\xi
_{Q^{\prime },k^{\prime }}]-\frac{1}{2},  \TCItag{26}
\end{eqnarray}

\bigskip

and from this we can write

\bigskip

\begin{multline}
\left\langle S_{j,m}^{A^{z}}(t)S_{j+R,n}^{A^{z}}(0)\right\rangle =<\left( 
\frac{1}{N}\dsum\limits_{Q,k,Q^{\prime },k^{\prime }}\exp \left(
-i(Q-Q^{\prime })dj\right) \times \right.  \notag \\
\left. \times \exp (i(E_{Qk}-E_{Q^{\prime }k^{\prime }})t)u_{Q,k,m}^{\ast
}u_{Q^{\prime },k^{\prime },m}\xi _{Q,k}^{\dag }\xi _{Q^{\prime },k^{\prime
}}-\frac{1}{2}\right) \times  \notag \\
\times \left( \frac{1}{N}\dsum\limits_{Q,k,Q^{\prime },k^{\prime }}\exp
\left( -i(Q-Q^{\prime })dl\right) u_{Q,k,n}^{\ast }u_{Q^{\prime },k^{\prime
},n}\xi _{Q,k}^{\dag }\xi _{Q^{\prime },k^{\prime }}-\frac{1}{2}\right) >. 
\tag{27}
\end{multline}

\bigskip

By using Wick's theorem we obtain

\bigskip

\begin{multline}
\left\langle S_{j,m}^{A^{z}}(t)S_{j+R,n}^{A^{z}}(0)\right\rangle =  \notag \\
\left( \frac{1}{N}\dsum\limits_{Q,k}\exp \left( -iQRdl\right) \exp
(iE_{Qk}t)u_{Q,k,m}^{\ast }u_{Q,,k,,m}\left\langle n_{Qk}\right\rangle
\right) \times  \notag \\
\left( \frac{1}{N}\dsum\limits_{Q,k}\exp \left( iQRdl\right) \exp
(-iE_{Qk}t)u_{Q,k,n}^{\ast }u_{Q,,k,,n}\left( 1-\left\langle
n_{Qk}\right\rangle \right) \right) +  \notag \\
\left( \frac{1}{N}\dsum\limits_{Qk}u_{Q,k,m}u_{Q,k,m}^{\ast }\left\langle
n_{Qk}\right\rangle \right) \times \left( \frac{1}{N}\dsum%
\limits_{Qk}u_{Q,k,n}u_{Q,k,n}^{\ast }\left\langle n_{Qk}\right\rangle
\right) +  \notag \\
-\frac{1}{2}\frac{1}{N}\dsum\limits_{Qk}u_{Q,k,m}u_{Q,k,m}^{\ast
}\left\langle n_{Qk}\right\rangle -\frac{1}{2}\frac{1}{N}\dsum%
\limits_{Qk}u_{Q,k,n}u_{Q,k,n}^{\ast }\left\langle n_{Qk}\right\rangle +%
\frac{1}{4},  \tag{28}
\end{multline}

\bigskip

which can be determined numerically.

We can also obtain the dynamic correlation between average cell spin
operators, $\tau _{l}^{z},$ defined as

\bigskip

\begin{equation}
\tau _{l}^{z}=\frac{1}{\left( n_{A}+n_{B}\right) }\left(
\dsum\limits_{m=1}^{n_{A}}S_{l,m}^{A^{z}}+\dsum%
\limits_{m=1}^{n_{B}}S_{l,m}^{B^{z}}\right) ,  \tag{29}
\end{equation}

\bigskip

which satisfy the equality $\left\langle \tau _{l}^{z}\right\rangle
=\left\langle S_{cel}^{z}\right\rangle .$ Therefore, by using eqs. (28) and
(29), we can determine the dynamic correlation function $\left\langle \tau
_{l}^{z}(t)\tau _{l+R}^{z}(0)\right\rangle $ by expressing it in terms of
dynamic correlations between site spins and it can be writen in the form

\bigskip

\begin{eqnarray}
\left\langle \tau _{l}^{z}(t)\tau _{l+R}^{z}(0)\right\rangle &=&\left[ \frac{%
1}{N\left( n_{A}+n_{B}\right) }\dsum\limits_{Q,k}\left( \left\langle
n_{Qk}\right\rangle -\frac{1}{2}\right) \right] ^{2}+  \notag \\
&&\left( \frac{1}{N\left( n_{A}+n_{B}\right) }\right) ^{2}\left(
\dsum\limits_{Q,k}\exp \left( -iQRdl\right) \exp (iE_{Qk}t)\left\langle
n_{Qk}\right\rangle \right) \times  \notag \\
&&\left( \dsum\limits_{Q,k}\exp \left( iQRdl\right) \exp (-iE_{Qk}t)\left(
1-\left\langle n_{Qk}\right\rangle \right) \right) .  \TCItag{30}
\end{eqnarray}

\bigskip

{\Large 4. The Longitudinal Dynamic Susceptibility}

The dynamic susceptibility $\chi _{Q}^{zz}(\omega )$ can be determined by
introducing the spatial and temporal Fourier transforms

\bigskip

\begin{equation}
\left\langle \tau _{Q}^{z}(t)\tau _{-Q}^{z}(0)\right\rangle =\sum_{R}\exp
(-iQR)\left\langle \tau _{l}^{z}(t)\tau _{l+R}^{z}(0)\right\rangle , 
\tag{31}
\end{equation}

\bigskip

and

\bigskip

\begin{equation}
\left\langle \tau _{Q}^{z}\tau _{-Q}^{z}\right\rangle _{\omega }=\frac{1}{%
2\pi }\int_{-\infty }^{\infty }\exp (i\omega t)\left\langle \tau
_{Q}^{z}(t)\tau _{-Q}^{z}(0)\right\rangle dt,  \tag{32}
\end{equation}

\bigskip

and by using eq. (30) we can write

\bigskip

\begin{eqnarray}
\left\langle \tau _{Q}^{z}\tau _{-Q}^{z}\right\rangle _{\omega } &=&N\delta
\left( \omega \right) \delta _{Q,0}\left[ \frac{1}{N\left(
n_{A}+n_{B}\right) }\dsum\limits_{Q_{1},k}\left( \left\langle
n_{Q_{1}k}\right\rangle -\frac{1}{2}\right) \right] ^{2}+  \notag \\
&&\frac{1}{N\left( n_{A}+n_{B}\right) ^{2}}\dsum\limits_{Q_{1},k}\delta
\left( \omega +E_{Qk}-E_{Q_{1}-Qk}\right) \left\langle
n_{Q_{1}k}\right\rangle \times  \notag \\
&&\times \left( 1-\left\langle n_{Q_{1}-Qk}\right\rangle \right) . 
\TCItag{33}
\end{eqnarray}

\bigskip

From this result we can obtain the dynamic susceptibility from the
expression [14]

\bigskip

\begin{equation}
\chi _{Q}^{zz}\left( \omega \right) =-2\pi <<\tau _{Q}^{z};\tau
_{-Q}^{z}>>=-\int_{-\infty }^{\infty }\frac{\left( 1-\exp (\beta \omega
)\right) \left\langle \tau _{Q}^{z}\tau _{-Q}^{z}\right\rangle _{\omega
^{^{\prime }}}d\omega ^{^{\prime }}}{\omega -\omega ^{^{\prime }}},  \tag{34}
\end{equation}

\bigskip

which can be written as

\bigskip

\begin{equation}
\chi _{Q}^{zz}\left( \omega \right) =\frac{1}{N\left( n_{A}+n_{B}\right) ^{2}%
}\dsum\limits_{Q_{1},k}\frac{\left\langle n_{Q_{1}k}\right\rangle
-\left\langle n_{Q_{1}-Qk}\right\rangle }{\omega +E_{Qk}-E_{Q_{1}-Qk}}. 
\tag{35}
\end{equation}

\bigskip

In the limit $\omega \rightarrow 0,Q\rightarrow 0$, we obtain the isothermal
susceptibility as in the uniform model [15].

The real and imaginary parts of $\chi _{Q}^{zz}\left( \omega \right) $ can
be determined from the previous expression by considering $\chi
_{Q}^{zz}\left( \omega -i\epsilon \right) $ in the limit $\epsilon
\longrightarrow 0,$ and from this we obtain explicitly the results

\bigskip

\begin{equation}
\func{Re}\chi _{Q}^{zz}\left( \omega \right) =\frac{1}{N\left(
n_{A}+n_{B}\right) ^{2}}\dsum\limits_{Q_{1},k}\frac{\left\langle
n_{Q_{1}k}\right\rangle -\left\langle n_{Q_{1}-Qk}\right\rangle }{\omega
+E_{Qk}-E_{Q_{1}-Qk}}  \tag{36}
\end{equation}

\bigskip

and

\bigskip

\begin{equation}
\func{Im}\chi _{Q}^{zz}\left( \omega \right) =\frac{\pi }{N\left(
n_{A}+n_{B}\right) ^{2}}\dsum\limits_{Q_{1},k}(\left\langle
n_{Q_{1}k}\right\rangle -\left\langle n_{Q_{1}-Qk}\right\rangle )\delta
(\omega +E_{Qk}-E_{Q_{1}-Qk}).  \tag{37}
\end{equation}

\bigskip

{\Large 5. Results and Conclusions}

The results for $J=1$, $J_{A}=2$, $J_{B}=3,$ $h_{A}=h_{B}=h$ and $%
n_{A}=n_{B}=2,$ at $T=0$ are shown in Figs. 2, 4, 5, 8, 10 and 12. For the
same interaction parameters and $n_{A}=2,n_{B}=3,$ the results are shown in
\ Figs. 3, 6, 7, 9, 11 and 13 $.$

In Figs. 2 and 3 we present the magnetization $\left\langle \tau
_{l}^{z}\right\rangle $ and the isothermal susceptibility $\chi _{T}^{zz}$
as functions of the field. In these cases, as expected, we have four and
five critical fields respectively, associated to the quantum transitions
induced by the field and the isothermal susceptibility diverges at these
points. In the first case ($n=n_{A}+n_{B},$ even) there is a plateau at zero
magnetization, whereas in the second one ($n=n_{A}+n_{B},$ odd) the
magnetization plateaus are at non-zero values. These results suggest that
these plateaus, which are associated to the gaps in the energy spectrum [7],
can also be related to the cluster spin states in a representation where the
effective spin $\tau _{l}^{z}$ is diagonal [10].

In Figs. 4-7 we present the real and imaginary parts of the dynamic
correlation function $\left\langle \tau _{l}^{z}(t)\tau
_{l+R}^{z}(0)\right\rangle ,$ as a function of time $t,$ for $R=1.$ For $%
n=4, $ Figs. 4 and 5, the results are presented for three different values
of the field, namely, $h=h_{jc}$ and $h=h_{jc}+\varepsilon $ ($\varepsilon
=0.001,$ $0.01),$ where $j=1,3$ respectively for each figure. For $n=5,$
Figs.6 and 7, the results are also presented for three different values of
the field, namely, $h=h_{jc}$ and $h=h_{jc}-\varepsilon $ ($\varepsilon
=0.001,$ $0.01), $ where $j=1,3$ respectively for each figure. In all cases,
the imaginary part of the correlation goes to zero as we approach the
magnetization plateaus, and the real part tends to the square of the
magnetization $\left\langle \tau _{l}^{z}\right\rangle .$ These results show
that within the plateaus, although $\left\langle
S_{j,m}^{A(B)^{z}}(t)S_{j+R,n}^{A(B)^{z}}(0)\right\rangle $ are not
constant, the field does not induce any quantum fluctuation on the
correlation $\left\langle \tau _{l}^{z}(t)\tau _{l+R}^{z}(0)\right\rangle .$
This means that the system, although not saturated, is in a frozen effective
spin state.

In Figs. 8 and 9 we show the static correlation function $\ \left\langle
\tau _{j}^{z}\text{ }\tau _{j+R}^{z}\right\rangle $ as a function of $R,$
the distance between cells, at $T=0$, and three different values of the
field. The correlation shows an oscillatory behaviour which is present for
any value of the field outside the plateaus, with increasing period as the
critical point is approached. On the other hand, at the critical field there
is no oscillatory behaviour, which means that the period of the oscillation
tends to infinity, and this result is still valid for any value of the field
within the plateaus. This is consistent with the scaling form and the
analytical continuation proposed for the two-spin correlation function $%
\left\langle S_{j}^{^{z}}(0)S_{ln}^{^{z}}(0)\right\rangle $ in the
homogeneous XY-model [16], where the correlation length is associated to the
oscillation period.

We have also verified that for $h\geq h_{4c}$ ($n=4,$ even) and $h\geq
h_{5c} $ ($n=5,$ odd)$,$ since the system is saturated, the dynamic
correlation function $<S_{jn}^{A(B)z}(t)S_{lm}^{A(B)z}(0)>$ does not depend
on time nor field, while in the intermediate plateaus it presents a time
dependence which is independent of the value of the field.

The static susceptibility $\chi _{Q}^{zz}(0),$ at $T=0,$ is shown in Figs.
10 and 11 for various wave-vectors and the two different cell sizes. As
already pointed out, the susceptibility $\chi _{0}^{zz}(0)$ is identical to
the isothermal one $\chi _{T}^{zz}$ and consequently diverges at the
critical points$.$ Also, independently of the wave-vector , the static
susceptibility goes to zero within the magnetization plateaus since in this
region the spacial modulation of the magnetic field does not induce any
fluctuation in the magnetization. The singularities present for the
different non-zero wave-vectors are related to the oscillations of the
correlation $\left\langle \tau _{l}^{z}(t)\tau _{l+R}^{z}(0)\right\rangle ,$
whereas the ones for $Q=0$ are related to the critical points. As pointed
out by Jullien and Pfeuty [17] for the homogeneous model and Lima and Gon\c{c%
}alves for the model on the supperlattice [18], the non-critical
singularities can also be associated to the unstable critical points.which
appear in the study of the system within the real space renormalization
approach.

Finally, in Figs. 12 and 13 we present the real and imaginary parts of the
dynamic susceptibility, $\chi _{q}^{zz}(\omega ),$ at $T=0$, for values of
the field close to critical ones and outside the plauteaus$,$ and the same
parameters of the previous figures. As expected, for \ a given wave-vector,
the bandwidth of the imaginary part of the response depends on the field
strength and on the size of the cell and there are no infinite
singularities.Also as expected, to the finite discontinuities at the band
edges of the imaginary part correspond divergences in the real part. It
should be noted that for values of the field within the plateaus the
response goes to zero for any non-zero wave-vector.

{\Large Acknowledgements}

The authors would like to thank the Brazilian agencies CNPq, Capes and Finep
for the partial financial support.

\begin{center}
\bigskip \pagebreak
\end{center}

\bigskip {\Large References}

[1] E. Lieb, T. Schultz, D. Mattis, Ann. Phys. (NY) \textbf{16} (1961) 407.

[2] J. Stolze, A. Nopert, G. M\"{u}ller, Phys. Rev. B \ \textbf{52} (1995)
4319

\ \ \ \ \ and references therein.

[3] V. M. Kontorovich, V. M. Tsurkenik, Sov. Phys. JEPT \textbf{26} (1968)
687.

[4] J. H. H. Perk, H. W. Capel, M. J. Zuilhof, Physica A \textbf{81} (1975)
319.

[5] J. H. H. Perk, Th. J. Siskens, H. W. Capel, Physica A \textbf{89} (1977)
304.

[6] J. P. de Lima, L.L. Gon\c{c}alves, J. Magn.\& Magn. Mater. \textbf{%
140-144},

\ \ \ \ \ 1606 (1995)

[7] J. P. de Lima, L. L. Gon\c{c}alves, J. Magn.\& Magn. Mater. \textbf{206}%
, (1999)

\ \ \ \ 135 and references therein.

[8] F. F. Barbosa Filho, J. P. de Lima, L. L. Gon\c{c}alves, J. Magn.\& Magn.

\ \ \ \ Mater. \textbf{226-230 }(2001) 638.

[9] O. Derzhko, J. Richter, O. Zaburannyi, Physica A \textbf{282} (2000) 495.

[10] O. Derzhko, J. Richter, O. Zaburannyi, J. Magn.\& Magn. Mater.

\ \ \ \ \ \ \textbf{222} (2000) 207.

[11] L. L. Gon\c{c}alves, \textit{Theory of Properties of Some
One-Dimensional Systems, }

\ \ \ \ \ \ \ (D.Phil.Thesis, University of Oxford, 1977).

[12] J. P. de Lima, L.L. Gon\c{c}alves, \textit{Mod. Phys. Lett.} \textbf{%
B22 }(1996) 1077.

[13] L. L. Gon\c{c}alves, J.P. de Lima, \textit{J. Phys.: Condens. Matter} 
\textbf{9} (1997) 3447.

[14] D. N. Zubarev, \textit{Sov. Phys. Usp. }\textbf{3} (1960) 320.

[15] S. Katsura, T. Horiguchi, M. Suzuki, Physica \textbf{46} (1970) 67.

[16] J. P. de Lima, L.L. Gon\c{c}alves, \textit{Mod. Phys. Lett.} \textbf{B
14\&15 }(1997) 871.

[17] R. Jullien, P. Pfeuty, Phys. Rev \textbf{B19} (1979) 4646.

[18] J. P. de Lima, L. L. Gon\c{c}alves, Mod. Phys. Lett. \textbf{B22}
(1996) 1077.

\pagebreak

{\Large Figure Captions}

\bigskip

\bigskip

\textbf{Fig.1}-Unit cell of the alternating superlattice.

\bigskip

\bigskip

\textbf{Fig. 2}- Average magnetization per unit cell (a) and the
susceptibility in the

\ \ \ \ \ \ \ \ \ \  field direction (b) as functions of the field, for $%
n_{A}=n_{B}=2,$ 

$\ \ \ \ \ \ \ \ \ \ \ J=1,$ $J_{A}=2,J_{B}=3,$\ and $h_{A}=h_{B}=h,$ at $T=0
$. The critical 

\ \ \ \ \ \ \ \ \ \ \ fields are $h_{c1}\cong 0.691,$\ $h_{c2}\cong 1.096,$\
and $\ h_{c3}\cong 1.5960.$

\bigskip

\bigskip

\bigskip

\textbf{Fig. 3}- Average magnetization per unit cell (a) and the
susceptibility in the 

\ \ \ \ \ \ \ \ \ \ \ field direction (b) as functions of the field, for $%
n_{A}=2,n_{B}=3,$ 

$\ \ \ \ \ \ \ \ \ \ \ J=1,$ $J_{A}=2,\ J_{B}=3,$ and $h_{A}=h_{B}=h,$ at $%
T=0$.\ The critical 

\ \ \ \ \ \ \ \ \ \ \ fields are $h_{c1}\cong 0.207,$\ $h_{c2}\cong 0.935,$ $%
h_{c3}\cong 1.207,$\ $h_{c4}\cong 2.161$ 

\ \ \ \ \ \ \ \ \ \ \ and $h_{c5}\cong 2.226.$

\bigskip

\bigskip

\bigskip

\textbf{Fig.4-} The real (a) and imaginary (b) parts of the correlation
function 

$\ \ \ \ \ \ \ \ \ \ \left\langle \tau _{j}^{z}(t)\tau
_{j+1}^{z}(0)\right\rangle ,$\ as functions of the time, at $T=0$, for $%
n_{A}=n_{B}=2,$ 

$\ \ \ \ \ \ \ \ \ \ J=1,$ $J_{A}=2,$\ $J_{B}=3,$ and $h=h_{1c}$ (plateau)$%
,h=h_{1c}+\varepsilon ,$ $\ $

\ \ \ \ \ \ \ \ \ \ with $\varepsilon =0.01$ and $0.001$.

\bigskip

\bigskip

\bigskip

\textbf{Fig.5}- The same as in Fig. 4 for $h=h_{3c}$ (plateaux)$,$ $%
h=h_{3c}+\varepsilon ,$ with

\ \ \ \ \ \ \ \ \ \ \ \ $\varepsilon =0.01$ and $0.001$.

\bigskip

\bigskip

\bigskip

\textbf{Fig.6-} The real (a) and imaginary (b) parts of the correlation
function 

$\ \ \ \ \ \ \ \ \ \ \left\langle \tau _{j}^{z}(t)\tau
_{j+1}^{z}(0)\right\rangle ,$\ as\ functions of the time, at $T=0$, for $%
n_{A}=2,n_{B}=3,$ 

$\ \ \ \ \ \ \ \ \ J=1,$ $J_{A}=2,$\ $J_{B}=3,$ and $h=h_{1c}$ (plateau)$%
,h=h_{1c}-\varepsilon ,$ 

$\ \ \ \ \ \ \ \ \ $with $\varepsilon =0.01$ and $0.001$.

\bigskip

\bigskip

\textbf{Fig.7}- The same as in Fig. 6) for $h=h_{3c}$ (plateaux)$,$ $%
h=h_{3c}-\varepsilon ,$ with

$\ \ \ \ \ \ \ \ \ \ \ \ \varepsilon =0.01$ and $0.001$.

\bigskip

\bigskip

\bigskip

\textbf{Fig.8}- The static correlation function $\ \left\langle \tau _{j}^{z}%
\text{ }\tau _{j+R}^{z}\right\rangle $ as a function of $R$ (distance

\ \ \ \ \ \ \ \ \ \ \ \ between cells), at $T=0$, for $n_{A}=n_{B}=2,$ $J=1,$
$J_{A}=2,$ $J_{B}=3,$

\ \ \ \ \ \ \ \ \ \ \ \ and for values of the field near and at the critical
field.

\bigskip

\bigskip

\bigskip

\textbf{Fig.9}- The static correlation function $\ \left\langle \tau _{j}^{z}%
\text{ }\tau _{j+R}^{z}\right\rangle $ as a function of $R$ (distance

\ \ \ \ \ \ \ \ \ \ \ \ between cells), at $T=0$, for $n_{A}=2,n_{B}=3,$ $%
J=1,$ $J_{A}=2,$ $J_{B}=3,$

\ \ \ \ \ \ \ \ \ \ \ and for values of the field near and at the critical
field.

\bigskip

\bigskip

\bigskip

\textbf{Fig.10}- Static susceptibility in the field direction, $\chi
_{q}^{zz}(0),$ at $T=0$, as a

\ \ \ \ \ \ \ \ \ \ \ \ function of the field for $n_{A}=n_{B}=2,$ $J=1,$ $%
J_{A}=2,$ $J_{B}=3,$ and

$\ \ \ \ \ \ \ \ \ \ \ \ h_{A}=h_{B}=h,$ and different values of $q$.

\bigskip

\bigskip

\bigskip

\textbf{Fig.11}- Static susceptibility in the field direction, $\chi
_{q}^{zz}(0),$ at $T=0$, as a

\ \ \ \ \ \ \ \ \ \ \ \ function of the field for $n_{A}=2,n_{B}=3,$ $J=1,$ $%
J_{A}=2,$ $J_{B}=3,$ and

$\ \ \ \ \ \ \ \ \ \ \ \ h_{A}=h_{B}=h,$ and different values of $q$.

\bigskip

\bigskip

\bigskip

\textbf{Fig.12}- The real (a) and imaginary parts of the dynamic
susceptibility in

\ \ \ \ \ \ \ \ \ \ \ \ the field\ direction, $\chi _{q}^{zz}(\omega ),$ at $%
T=0$, as a function of frequency for,

\ \ \ \ \ \ \ \ \ \ \  $n_{A}=n_{B}=2,J=1,$ $\ J_{A}=2,$ $J_{B}=3,$ and $%
h=0.693$.

\bigskip

\bigskip

\bigskip

\textbf{Fig.13}- The real (a) and imaginary parts of the dynamic
susceptibility in 

\ \ \ \ \ \ \ \ \ \ \ \ the field\ direction, $\chi _{q}^{zz}(\omega ),$ at $%
T=0$, as a function of frequency for, 

$\ \ \ \ \ \ \ \ \ \ \ n_{A}=2,n_{B}=3,\ J=1,$ $\ J_{A}=2,$ $J_{B}=3,$ and $%
h=0.2$.

\bigskip 

\end{document}